\documentclass[onecollarge,natbib]{svjour2}
\bibpunct{[}{]}{;}{n}{}{,} 
\smartqed  
\usepackage{graphicx}
\journalname{Few-Body Systems (EFB22)}
\usepackage{capt-of}
\usepackage{amsmath}

\usepackage{amssymb}
\usepackage{relsize}

\usepackage{xcolor}

\newcommand{\bra}{\langle}
\newcommand{\ket}{\rangle}
\newcommand{\bs}[1]{\ensuremath{\boldsymbol{#1}}}

\newcommand{\be}{\begin{equation}}
\newcommand{\ee}{\end{equation}}
\newcommand{\bea}{\begin{eqnarray}}
\newcommand{\eea}{\end{eqnarray}}

\DeclareMathOperator*{\SumInt}{%
\mathchoice%
  {\ooalign{$\displaystyle\sum$\cr\hidewidth$\displaystyle\int$\hidewidth\cr}}
  {\ooalign{\raisebox{.14\height}{\scalebox{.7}{$\textstyle\sum$}}\cr\hidewidth$\textstyle\int$\hidewidth\cr}}   
  {\ooalign{\raisebox{.2\height}{\scalebox{.6}{$\scriptstyle\sum$}}\cr$\scriptstyle\int$\cr}}
  {\ooalign{\raisebox{.2\height}{\scalebox{.6}{$\scriptstyle\sum$}}\cr$\scriptstyle\int$\cr}}  
}

\begin{document}

\title{
Nuclear polarization effects in muonic atoms
~\thanks{This work is supported in parts by the Natural Sciences
  and Engineering Research Council (NSERC), the National Research
  Council of Canada, and the Israel Science Foundation (Grant number
  954/09).}
}


\author{C.~Ji         \and
        N.~Nevo Dinur  \and
        S.~Bacca     \and
        N.~Barnea
}

\authorrunning{Chen Ji \textit{et al.}} 

\institute{C.~Ji \at
              TRIUMF, 4004 Wesbrook Mall, Vancouver, BC V6T 2A3, Canada\\
              \email{jichen@triumf.ca}            
\and
           N.~Nevo Dinur  \at
              Racah Institute of Physics, The Hebrew University, Jerusalem 91904, Israel\\
           \email{nir.nevo@mail.huji.ac.il}
\and
           S.~Bacca \at
              TRIUMF, 4004 Wesbrook Mall, Vancouver, BC, V6T 2A3, Canada\\  
              Department of Physics and Astronomy, University of Manitoba, Winnipeg, MB, R3T 2N2, Canada \\
              \email{bacca@triumf.ca}
\and
              N. Barnea \at
              Racah Institute of Physics, The Hebrew University of Jerusalem, 91904, Jerusalem, Israel\\
              Tel.: 972-2-6585138\\
              Fax: 972-2-6584437\\
              \email{nir@phys.huji.ac.il}
}

\date{Received: date / Accepted: date}

\maketitle

\begin{abstract}
We illustrate how nuclear polarization corrections in muonic atoms can be formally connected to inelastic
response functions of a nucleus. We first discuss the point-nucleon approximation and then include finite-nucleon-size corrections.
As an example, we compare our {\it ab-initio} calculation of the third Zemach moment in  $\mu\,^4$He$^+$ to previous phenomenological results.  
\end{abstract}


\section{Introduction}
Recent measurements of the $\mu\,$H Lamb shift $\Delta E$ ($2S$-$2P$
transition) at PSI~\cite{Pohl:2010zza,Antognini:1900ns}
allowed for an unprecedented precise determination of the proton charge radius, which deviates, however, 
by $7 \sigma$ from the CODATA value based on $e\,$H measurements~\cite{Mohr:2012tt}.  
This discrepancy challenges our understanding of experimental errors and theoretical calculations. 
To investigate this discrepancy, spectroscopy measurements in other muonic atoms,
e.g.,
$\mu\,$D, $\mu\,^3$He$^+$ and $\mu\,^4$He$^+$, are planned at PSI~\cite{Antognini:2011zz}. 
The extraction of the nuclear charge radius
$\bra R^2_c\ket^{1/2}$ from a $\Delta E$ measurement,
based on the expression
\begin{equation} \label{eq:E2s2p}
\Delta E \equiv \delta_{QED}+\delta_{pol} + \delta_{Zem} 
                + m_r^3 (Z\alpha)^4 \bra R^2_c\ket/12. 
\end{equation}
must be accompanied by theoretical estimates of the QED corrections $\delta_{QED}$,
as well as the Zemach moment $\delta_{Zem}$
and the nuclear polarization $\delta_{pol}$, which are the elastic and inelastic nuclear structure corrections, respectively.  
The 
Zemach contribution is defined via the nuclear charge density $\tilde{\rho}_0(\bs{R})$ as 
\begin{equation}
\label{eq:zemach3}
\delta_{Zem} = - \frac{m_r^4}{24}(Z\alpha)^5 \iint d \bs{R} d \bs{R}' \left|\bs{R}-\bs{R}'\right|^3 \tilde{\rho}_0(\bs{R})\tilde{\rho}_0(\bs{R}').
\end{equation}
In the non-relativistic limit, the general Hamiltonian for a muonic atom can be expressed by
\begin{equation}
\label{eq:H-muHe}
   H = H_{nucl} + H_{\mu } -\Delta H \,,
\end{equation}
where $H_{nucl}$ is the nuclear Hamiltonian, $H_{\mu}= p^2/2m_r-Z\alpha/r$  the muonic Hamiltonian,  $m_r$ the reduced mass
and $\Delta H$  the difference between the muonic Coulomb interaction with the point-nucleus and
the sum of the interactions with each of the $Z$
 protons, located  at a distance $\bs{R}_a$  from the  center of mass:
\begin{equation}\label{eq:dH}
\Delta H = \sum_a^Z \Delta V(\bs{r},\bs{R}_a)\equiv 
         \sum_a^{Z} \alpha \left(\frac{1}{\left|\bs{r}-\bs{R}_a\right|} -\frac{1}{r}\right).
\end{equation}
Nuclear structure corrections are calculated using  second-order perturbation theory on
$\Delta H$.
Since a muon 
interacts more closely with the nucleus
than does an electron,  muonic atoms are very sensitive to nuclear corrections~\cite{Borie:1982ax, Borie:2012zz}. 
Calculations in $\mu\,$D and $\mu\,^4$He$^+$ have demonstrated the importance of  using state-of-the-art methods and nuclear Hamiltonians 
to predict such corrections at a percentage accuracy~\cite{Ji:2013oba,Leidemann:1994vq,Pachucki:2011xr,Friar:2013rha}.
Here, we  bridge the nuclear effects to inelastic response functions, and to charge  and transition densities.
Then, we present results on the Zemach moment in $\mu\,^4$He$^+$ (see also~\cite{Ji:2013oba}).

\section{Polarization in the point-nucleon approximation}

From second order perturbation theory, the nuclear polarization $\delta^{A}_{pol}$ becomes\footnote{The upper index $A$ indicates that the
intrinsic nucleon polarization is not included, see~\cite{Ji:2013oba} for more details.}
\begin{equation}\label{eq:dE-pol}
  \delta^{A}_{pol} = -\SumInt \limits_{N\neq N_0} \!\!\iint d\bs{R}d\bs{R}'
     \rho_N^*(\bs{R})P(\bs{R},\bs{R}',\omega_N) \rho_N(\bs{R}'),
\end{equation}
where
$\rho_N(\bs{R}) = \bra N | \hat{\rho}(\bs{R}) |N_0\ket$ 
is the matrix element  of the  point-nucleon charge density operator  
$\hat{\rho}(\bs{R}) \equiv  \sum_{a}^{Z} \delta(\bs{R}-\bs{R}_a)$, $\omega_N=E_N-E_{N_0}$, and $E_{N_0}$, $E_{N}$, $|N_0\ket$ and $|N\ket$
are the nuclear ground- and excited-state energies and wave-functions,
respectively. The symbol $\SumInt$ indicates a sum over discrete and an integration 
 over
continuum states. The nucleus is excited into all possible
states but $E_{N_0}$, thus $ \delta^{A}_{pol} $ is an inelastic contribution.
The muonic matrix element $ P(\bs{R},\,\bs{R}', \omega_N)\equiv P$ is given by
\begin{align}\label{eq:P-NR}
P = -Z^2 \int d\bs{r} d\bs{r}'
     \Delta V(\bs{r},\bs{R}) \bra \mu_0|\bs{r}\ket
     \bra \bs{r}
       |\frac{1}{H_\mu +\omega_N-\epsilon_{\mu_0}} | \bs{r}'\ket \bra \bs{r}'
       |\mu_0\ket \Delta V(\bs{r}',\bs{R}'),
\end{align}
where
$\epsilon_{\mu_0}$ and $|\mu_0\ket$ are 
the unperturbed muon atomic energy and wave-function in either the 2$S$ or
2$P$ states. 
The leading  contribution to $\delta^{A}_{pol}$ is
obtained by
neglecting in Eq.~\eqref{eq:P-NR} the Coulomb-potential part of $H_\mu$.
Considering only contributions to the 2$S$ state  we get
\begin{align}
\label{eq:PNR}
P=&
-Z^2 \phi^2(0) \int \frac{d\bs q}{(2\pi)^3}
\left(\frac{4\pi\alpha}{q^2}\right)^2 \left(1-e^{i\bs q\cdot \bs R}\right) 
\frac{1}{q^2/2 m_r +\omega_N} \left(1-e^{-i\bs q\cdot \bs{R}'}\right),
\end{align}
where $\phi^2(0)=(m_rZ\alpha)^3 /8\pi$ is the $\mu$ wave function at the origin.
Since  $\bs q$ can be large in the integral, we should not expand the plane wave in multipoles.
Instead, we first integrate over $\bs q$ in Eq.~\eqref{eq:PNR}, obtaining
\begin{equation}
\label{eq:PNR-full}
P = -\frac{2\pi\alpha^2 Z^2 \phi^2(0)}{m_r\omega_N^2} 
\frac{1}{|\bs{R}-\bs{R}'|} \left[ e^{-\sqrt{2m_r
      \omega_N}|\bs{R}-\bs{R}'|} -1+\sqrt{2m_r
    \omega_N}|\bs{R}-\bs{R}'| - m_r \omega_N |\bs{R}-\bs{R}'|^2\right]. 
\end{equation}
The quantity $\left|\bs{R}-\bs{R}'\right|$ indicates the ``virtual'' distance a proton travels during the two-photon exchange. Due to uncertainty principle it becomes $\left|\bs{R}-\bs{R}'\right| \sim 1/\sqrt{2M_A \omega_N}$. 
Therefore, $\left|\bs{R}-\bs{R}'\right|\sqrt{2m_r \omega_N}$
\newpage
\noindent
$\sim \sqrt{m_r/M_A}$ can be used as a systematic expansion parameter.
Up to 4$^{th}$ order this expansion yields
\begin{equation}
\label{eq:PNR-exp}
P(\left|\bs{R}-\bs{R}'\right|,\omega_N) \simeq   \frac{m_r^3 (Z\alpha)^5}{12} 
\sqrt{\frac{2m_r}{\omega_N}} 
\left[\left|\bs{R}-\bs{R}'\right|^2 - \frac{\sqrt{2 m_r \omega_N}}{4} \left|\bs{R}-\bs{R}'\right|^3
+ \frac{m_r \omega_N}{10} \left|\bs{R}-\bs{R}'\right|^4 \right].
\end{equation}
The terms in Eq.~\eqref{eq:PNR-exp} contribute to $\delta^{A}_{pol}$ at different orders in the $\sqrt{m_r/M_A}$ expansion,
and are evaluated using nuclear response functions defined as
\begin{equation}
S_{O}(\omega) \equiv  \frac{1}{2J_0+1} \!\!\!\!
 \SumInt \limits_{N\neq N_0, J} \!\!\!\!
 |\bra N_0 J_0 ||\hat O ||N J\ket|^2
\delta(\omega-\omega_N),
\end{equation}
where $\hat O$ is a general operator and $J_0$ ($J$) is the angular momentum of the ground (excited) state.
The leading $\left|\bs{R}-\bs{R}'\right|^2$ term in Eq.~\eqref{eq:PNR-exp} is related to the electric-dipole excitation and  yields
\begin{equation}
\label{eq:del-R2}
\delta^{(0)}_{D1}  =-\frac{2\pi m_r^3}{9}  (Z\alpha)^5 \int^\infty_{\omega_{\rm th}}
d\omega \sqrt{\frac{2m_r}{\omega}} S_{D_1}(\omega), 
\end{equation}
where $\hat D_1 = \frac{1}{Z} \sum_a^Z R_a Y_1(\hat R_a)$, $Y_1$ is the
rank-1 spherical harmonics, and $\omega_{\rm th}$ 
denotes the lowest excitation energy of the nucleus.
The sub-leading
$\left|\bs{R}-\bs{R}'\right|^3$ term is independent of $\omega$, which allows substituting 
$\SumInt_{N\neq N_0} |N\ket \bra N|$ with $1-|N_0\ket \bra N_0|$, and relating this inelastic term to two elastic ones 
\begin{eqnarray}
\label{eq:del-R3}
\delta^{(1)}=\delta^{(1)}_{R3pp}+\delta^{(1)}_{Z3}=
   &-&\frac{m_r^4}{24}(Z\alpha)^5\iint d \bs{R} d \bs{R}' |\bs{R}-\bs{R}'|^3 
\bra N_0| \hat{\rho}^\dagger(\bs{R})\hat{\rho}(\bs{R}')| N_0\ket \\
\nonumber
   &+&\frac{m_r^4}{24}(Z\alpha)^5\iint d \bs{R} d \bs{R}' |\bs{R}-\bs{R}'|^3  \rho_0(\bs{R})\rho_0(\bs{R}')
\end{eqnarray}
where $\rho_0(\bs{R})\equiv \rho_{N_0}(\bs{R}) = \bra N_0 | \hat{\rho}(\bs{R})| N_0\ket$ is the
nuclear charge density distribution in the point-nucleon approximation.
The first term in Eq.~\eqref{eq:del-R3},  $\delta^{(1)}_{R3pp}$,
is the 3$^{rd}$ moment of the proton charge correlation function.
The term $\delta^{(1)}_{Z3}$
is the 3$^{rd}$ Zemach moment~\cite{Friar:1997tr, Pachucki:2011xr, Friar:2013rha}, 
which cancels the elastic contribution $\delta_{Zem}$ in
Eq.~(\ref{eq:zemach3}), when $\tilde{\rho}_0(\bs{R})$ is defined in the point-nucleon limit as $\rho_0(\bs{R})$. 
Finally, contributions from the sub-sub-leading $\left|\bs{R}-\bs{R}'\right|^4$
term 
are
\begin{align}
\label{eq:del-R4}
\delta^{(2)} &= \frac{m_r^5}{18} (Z \alpha)^5  \int^\infty_{\omega_{\rm th}}
d\omega \sqrt{\frac{\omega}{2m_r}} 
\left[ S_{R^2}(\omega)
+ \frac{16\pi}{25} S_{Q}(\omega)
+ \frac{16\pi}{5}  \mathcal{S}_{D_1 D_3}(\omega) \right],
\end{align}
  where $S_{R^2}$ and $S_{Q}$ are, respectively,
the
  monopole $\hat R^2 = \frac{1}{Z} \sum_a^Z R^2_a$ and quadrupole
  \mbox{$\hat Q = \frac{1}{Z} \sum_a^Z R^2_a  Y_2(\hat R_a)$} structure functions.
$\mathcal{S}_{D_1D_3}$ indicates the interference between
 $\hat D_1$ and $\hat D_3 =\frac{1}{Z} \sum_a^Z R^3_a
Y_1(\hat R_a)$~\cite{Ji:2013oba}.

\section{Corrections due to the finite nucleon charge distributions}
Considering the finite charge distributions of the nucleons, the proton position in 
Eq.~\eqref{eq:dH} should be replaced by a convolution over the proton 
charge density, and similarly  for the neutron. 
Therefore, the point-proton muonic matrix element $P$ in Eq. ~\eqref{eq:PNR}
is replaced by three finite-size terms: 
 proton-proton $P_{pp}$, neutron-proton $P_{np}$, and neutron-neutron
 $P_{nn}$.
The muonic matrix elements $P_{pp}$ and $P_{np}$ are 
\begin{align}
\label{eq:P-pp}
 P_{pp}=&
-Z^2 \phi^2(0) \int \frac{d\bs q}{(2\pi)^3}
\left(\frac{4\pi\alpha}{q^2}\right)^2 \left[1-G^E_p(q^2)e^{i\bs q\cdot \bs R}\right] 
\frac{1}{q^2/2 m_r +\omega} \left[1-G^E_p(q^2)e^{-i\bs q\cdot \bs{R}'}\right],
\\
\label{eq:P-np}
 P_{np}=&\,
2NZ \phi^2(0) \int \frac{d\bs q}{(2\pi)^3}
\left(\frac{4\pi\alpha}{q^2}\right)^2 \left[1-G^E_p(q^2)e^{i\bs q\cdot \bs R}\right]
\frac{1}{q^2/2 m_r +\omega}\, G^E_n(q^2)e^{-i\bs q\cdot \bs{R}'},
\end{align}
and we neglect the neutron-neutron term. $G^E_p$ and $G^E_n$ are
respectively the proton and neutron charge form factors. Here, we adopt the
form factors defined in Ref.~\cite{Friar:2005je}:
$G^E_p(q^2)=(1+q^2/\beta^2)^{-2}$ and $G^E_n(q^2)=\lambda
q^2(1+q^2/\beta^2)^{-3}$, where $\beta$ and $\lambda$ are fixed by the proton
and neutron charge radii via $\beta=\sqrt{12} /\bra r_p^2 \ket^{1/2}$ and
$\lambda=-\bra r_n^2 \ket/6$. Similar to solving Eq.~\eqref{eq:PNR}, we
integrate over $\bs q$ in Eq.~\eqref{eq:P-pp} and \eqref{eq:P-np}, and extract
the dominant nucleon-size corrections. 
For $^4$He the calculation of the nucleon-size corrections is further simplified, since $N=Z$ and $\rho^{(p)}_{0}(\bs R)=\rho^{(n)}_{0}(\bs R)= \rho_{0}(\bs R)$ assuming isospin symmetry in the ground-state. Because corrections to $\delta^{(0)}$ vanish, the leading nucleon-size (NS) correction is
\begin{eqnarray}\label{eq:delta_NS}
  \delta^{(1)}_{NS}=\delta^{(1)}_{R1pp}+\delta^{(1)}_{Z1}=
   & - & m_r^4(Z \alpha )^5 \left[\frac{2}{\beta^2}-\lambda \right]
       \iint d\bs{R}d\bs{R}' |\bs{R}-\bs{R}'|
       \bra N_0| \hat{\rho}^\dagger(\bs{R})\hat{\rho}(\bs{R}')| N_0\ket\\
\nonumber
   & +& m_r^4(Z \alpha )^5 \left[\frac{2}{\beta^2}-\lambda \right]
       \iint d\bs{R}d\bs{R}' |\bs{R}-\bs{R}'|
         \rho_0(\bs{R})\rho_0(\bs{R}')\,.
\end{eqnarray}
Here, similarly to $\delta^{(1)}$, contributions to $\delta^{(1)}_{NS}$ 
are divided into two terms, $\delta^{(1)}_{R1pp}$ and $\delta^{(1)}_{Z1}$.
The latter one is the first Zemach moment and is the leading finite-nucleon-size correction to the elastic
nuclear structure correction.
In fact, the combination of $\delta^{(1)}_{Z3}$ and $\delta^{(1)}_{Z1}$ cancels out the elastic Zemach term $\delta_{Zem}$
in Eq.~(\ref{eq:zemach3}) at the order discussed here.

\section{Results: The Zemach elastic term }

 In Ref.~\cite{Borie:2012zz}, $\delta_{Zem}$ for $\mu\,^4$He$^+$ is
 approximated as $-1.40(4) \bra R_c^2 \ket^{3/2}$ using
 the continuous charge distribution approximation in
 Ref.~\cite{Friar:1979zz}, with the error representing the model dependence.
 Using $R_c=1.681$ fm determined from $e$-$^4$He scattering data~\cite{Sick:2008zza},
 one obtains $\delta_{Zem} = -6.65(19)$ meV
 with this phenomenological model. 
We calculated $\delta^{(1)}_{Z3}$ and $\delta^{(1)}_{Z1}$ in
Ref.~\cite{Ji:2013oba} using state-of-the-art nuclear potentials, {i.e.,}
AV18/UIX potentials and EFT N$^3$LO ($NN$) /N$^2$LO ($NNN$) potentials. In Ref.~\cite{Ji:2013oba},
$\beta$ corresponds to
$\bra r_p^2 \ket^{1/2} = 0.8409$ fm
from $\mu\,$H Lamb shift measurements~\cite{Antognini:1900ns}. 
To compare with
$\delta_{Zem}$ based on $e$-$^4$He scattering data, for consistency we
should use the value of $\beta$ corresponding to $\bra r_p^2\ket^{1/2}$ from
$e\,$H data. Using $\bra r_p^2\ket^{1/2} =0.8775$ fm~\cite{Mohr:2012tt}, we
calculated $\delta^{(1)}_{Z3}+\delta^{(1)}_{Z1}$ and obtained 6.12 meV for
the AV18/UIX potential and 6.53 meV for 
the EFT potential.
Averaging these two numbers, we 
obtain $\delta^{(1)}_{Z3}+\delta^{(1)}_{Z1} = 6.32(21)$ meV. This is
consistent with the phenomenological value of $\delta_{Zem}$ mentioned above.

\section{Conclusions}
We illustrate the calculation of the nuclear corrections to the Lamb shift in muonic atoms, which plays an essential role in the accurate extraction of nuclear charge radii from  spectroscopy measurements.
We show that the non-relativistic nuclear polarization effects can be systematically expanded in powers of
$\left|\bs{R}-\bs{R}'\right|\sqrt{2m_r \omega}$.
We also discuss finite-nucleon-size corrections.
Combining the point-nucleon results with the finite-nucleon-size corrections,
we provide an {\it ab-initio} calculation of the $3^{\rm rd}$ Zemach moment in $\mu\,^4$He$^+$,
which agrees with previous phenomenological calculations.


\end{document}